\documentclass[RNAAS]{aastex62}

\newcommand\sands{S\&S}

\usepackage{url}
\usepackage{amsmath} % \begin{cases} does not work without this, even though it's supposed to be in
\begin{document}

\title{A Search for Refraction in {\it Kepler} Photometry of Gas Giants}

%% Note that the corresponding author command and emails has to come
%% before everything else. Also place all the emails in the \email
%% command instead of using multiple \email calls.

\correspondingauthor{Holly A. Sheets}
\email{holly.sheets@mail.mcgill.ca}

\author{Holly A. Sheets}
\altaffiliation{Trottier Postdoctoral Fellow}
\affiliation{Department of Physics, McGill University, \\ 3600 rue University, Montreal, QC, H3A 2T8, Canada}
\affiliation{McGill Space Institute, McGill University, \\ 3550 rue University, Montreal, QC H3A 2A7}
\affiliation{Institut de Recherche sur les Exoplan\`etes, D\'epartement de physique, Universit\'e de Montr\'eal, \\ 2900 boul. \'Edouard-Montpetit, Montr\'eal, Qc, H3T 1J4, Canada}
\affiliation{NASA Astrobiology Institute's Virtual Planetary Laboratory, USA}

\author{Laurent Jacob}
\altaffiliation{Trottier Summer Intern}
\affiliation{Institut de Recherche sur les Exoplan\`etes, D\'epartement de physique, Universit\'e de Montr\'eal, \\ 2900 boul. \'Edouard-Montpetit, Montr\'eal, Qc, H3T 1J4, Canada}

%%% The \author command can take an optional ORCID.
\author{Nicolas B. Cowan}
\affiliation{Department of Physics, McGill University, \\ 3600 rue University, Montreal, QC, H3A 2T8, Canada}
\affiliation{Department of Earth \& Planetary Sciences, McGill University, \\ 3450 rue University, Montreal, QC, H3A 0E8, Canada}
\affiliation{McGill Space Institute, McGill University, \\ 3550 rue University, Montreal, QC H3A 2A7}
\affiliation{Institut de Recherche sur les Exoplan\`etes, D\'epartement de physique, Universit\'e de Montr\'eal, \\ 2900 boul. \'Edouard-Montpetit, Montr\'eal, Qc, H3T 1J4, Canada}

\author{Drake Deming}
\affiliation{NASA Astrobiology Institute's Virtual Planetary Laboratory, USA}
\affiliation{Department of Astronomy, University of Maryland, College Park, MD 20742-2421}


\section{}   
Refraction can lead to a brightening just before ingress and just after egress of a transit, as light passes through the exoplanet's atmosphere and is refracted into our line of sight \citep{sands,mmc,mm,dalba,alp}. Refraction just outside of transit has been seen and modeled in our own solar system during transits of Venus \citep{pasachoff,tanga,gm_venus}. 
For short-period planets, the model of \citet[][hereafter \sands]{sands} implies refraction peaks typically under 100 parts per million (ppm) and comparable in duration to ingress and egress.  {\it Kepler} photometry \citep{borucki} currently provides the best opportunity for detecting refraction. We search for the signature of refraction just outside of transit in {\it Kepler} photometry of 45 gas giants and firmly rule out the \sands~model for four candidates. 

We select {\it Kepler} Objects of Interest (KOIs) with radii at least twice that of Earth for which the \sands~Equation (30) implies a peak effect greater than 10 parts per million (ppm), adjusted for Rayleigh scattering using their Equations (40)-(45).  We eliminate KOIs with grazing transits as well as those identified in \citet{ford}, \citet{mazeh}, and \citet{holczer} as having significant transit timing variations.  We also eliminate a few KOIs identified by \citet{holczer} as likely planetary false positives based on the behavior of the light curves, leaving 45 planet candidates.  To calculate the expected effect, we adopt the masses predicted in \citet{chen_koi_mass}.

We use the simple aperture photometry fluxes and apply corrections for crowding and for target flux missed by the optimal aperture \citep{kam}.  We perform a 3-$\sigma$ clip using a median filter over 101 points with two iterations.  To normalize each transit, we take a light curve segment seven times the transit duration and centered on the mid-transit time $t_0$, mask the full transit and half the duration before and after, and fit a first- to fifth-order polynomial.  The whole segment is then divided by the polynomial that gave the best fit.

We phase fold the normalized data from the masked region above for each candidate, binning in 5-minute increments, and fit our refraction model.  We use {\tt batman} \citep{batman} to model the transit and add refraction shoulders:
\begin{equation}
\text{shoulder} = 
  \begin{cases} 
    h_1\exp(c(t-t_1)) & t < t_0 \\
    h_4\exp(c(t_4-t)) & t > t_0,
  \end{cases}
\end{equation}
where $h_1$ and $h_4$ are the scale of the shoulder pre- and post-transit, $t_1$ is the time of first contact, and $t_4$ is the time of last contact. The constant $c = \ln(100)/(0.25d)$, where $d$ is the duration of the transit, is chosen such that the refraction shoulder falls to 1\% of the maximum by a quarter of the transit duration from first/last contact, in agreement with \citet{mm}.  We model each shoulder separately, because clouds and hazes may behave differently at the east and west terminators \citep[e.g.][]{k7b,kempton}. We fit the data with {\tt emcee} \citep{emcee}, adopting uniform priors from 0 to 500 ppm for the shoulder heights and from -100 to 100 ppm for a baseline offset for the out-of-transit light curve. We also allow the scaled planet radius $R_p/R_s$ and the scaled semi-major axis $a/R_s$ to vary, with gaussian priors based on the values from the Data Release 25 KOI table \citep{dr25} downloaded from the NASA Exoplanet Archive.  We fit for the limb darkening parameters following \citet{limb}.  

\begin{figure}[h!]
\begin{center}
\includegraphics[scale=0.95,angle=0]{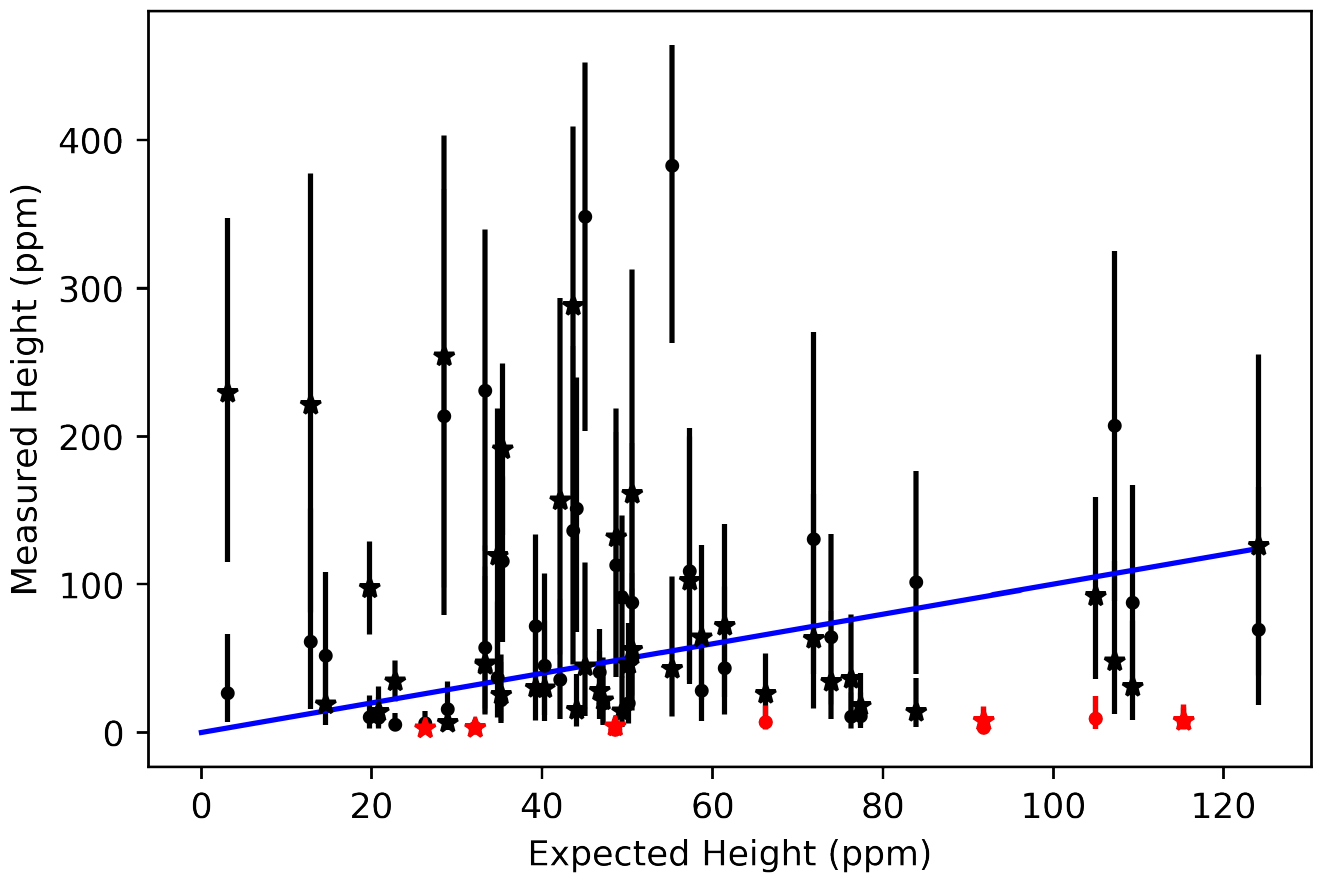}
\caption{The expected heights versus the measured heights with uncertainties for each candidate.  Pre-transit heights are plotted with circles, and post-transit with stars. Red highlights heights more than 5-$\sigma$ below the expected height.  In blue is the 1:1 line.  \label{fig:1}}
\end{center}
\end{figure}

Figure \ref{fig:1} shows the measured shoulder heights and their uncertainties, plotting the pre- and post-transit shoulder for each candidate separately, versus the expected heights.  Four systems (KOIs 108.01, 144.01, 161.01, 261.01) have both shoulders at least 5-$\sigma$ below their expected heights, while two have the pre-transit shoulder (KOIs 197.01, 1860.01) and one has the post-transit shoulder (KOI 281.01) measured at least 5-$\sigma$ below the expected value. 

As an independent test of the significance of the refraction shoulders, we repeat the fit at quadrature as a control.  We compare the transit distribution, including both the pre- and post-transit shoulder for each candidate, to that at quadrature using a K-S test.  The quadrature distribution includes both shoulder heights at both phase 0.25 and phase 0.75.  We find that there is no significant difference (P = 0.417) between the transit distribution and the quadrature distribution.

%\section{Conclusions}
We have not detected a refraction signal in any individual planet, nor in the ensemble.  \citet{dalba} and \citet{alp} predict that the refraction signal for our sample would be far weaker than expected from the \sands~model.  \citet{alp} also searched the {\it Kepler} photometry for a sample at longer periods than our sample and found no sign of refraction.  

%% An example figure call using \includegraphics

\end{document}